\documentclass[11pt,a4paper]{article}
\pdfoutput=1

\usepackage{jcappub}

\usepackage{mathrsfs}
\usepackage{bbm}
\usepackage{bm}
\usepackage{dsfont}
\usepackage{tensor}
\usepackage{xspace}
\usepackage{lmodern}
\usepackage{wasysym}
\usepackage{microtype}
\usepackage{empheq}
\usepackage{ifthen}
\usepackage{amsmath}
\usepackage{wrapfig}
\usepackage{cleveref}
\usepackage{xcolor}
\usepackage{siunitx}
\DeclareSIUnit\parsec{pc}
\usepackage{floatrow}
\usepackage[normalem]{ulem}
\newcommand\rd{{\rm d}}
\newcommand\balpha{\bm{\alpha}}
\newcommand\btheta{\bm{\theta}}

\newcommand\bbeta{\bm{\beta}}

\newcommand\bx{\bm{x}}

\newcommand\ri{{\rm i}}

\newcommand\bA{\bm{\mathcal{A}}}

\newcommand\thetaE{\theta_{\mathrm{E}}}
\newcommand\tthetaE{\tilde{\theta}_{\mathrm{E}}}
\newcommand\ttheta{\tilde{\theta}}
\newcommand\os{\mathrm{os}}
\newcommand\ds{\mathrm{ds}}
\newcommand\od{\mathrm{od}}

\newcommand\vect[1]{\boldsymbol{#1}}

\newcommand{\uvect}[1]{\hat{\vect{#1}}}

\newcommand\e[1]{_{\mathrm{#1}}}
\newcommand\h[1]{^{\mathrm{#1}}}

\newcommand{\delimiters}[4][]{
\ifthenelse{ \equal{#1}{1} }{  #2 #3 #4  }
					{ \ifthenelse{\equal{#1}{2}}{ \big#2 #3 \big#4 }
						{ \ifthenelse{\equal{#1}{3}}{ \Big#2 #3 \Big#4 }
							{ \ifthenelse{\equal{#1}{4}}{ \bigg#2 #3 \bigg#4 }
								{ \ifthenelse{\equal{#1}{5}}{ \Bigg#2 #3 \Bigg#4 }
									{ \left#2 #3 \right#4 }
								}
							}
						}
					}
													}

\definecolor{blue4}{RGB}{0,0,143}
\definecolor{red4}{RGB}{143,0,0}
\definecolor{orange}{RGB}{255,128,0}
\definecolor{darkcyan}{RGB}{0,128,128}
\definecolor{olive}{RGB}{0,128,0}
\definecolor{purple}{RGB}{128,0,128}
\definecolor{cyan2}{RGB}{0,255,255}
\definecolor{fushia}{RGB}{255,0,255}
\definecolor{mygray}{gray}{0.5}
\definecolor{lightgray}{gray}{0.85}

\makeatletter
\def\@fpheader{\relax}
\makeatother

\title{Line-of-sight effects on double source plane lenses}

\author[a]{Daniel Johnson,}

\affiliation[a]{Laboratoire Univers et Particules de Montpellier (LUPM), 
CNRS \& Université de Montpellier (UMR-5299),
Parvis Alexander Grothendieck, F-34095 Montpellier Cedex 05, France}

\emailAdd{daniel.johnson@umontpellier.fr}

\author[b]{Thomas Collett,}

\affiliation[b]{Institute of Cosmology and Gravitation, University of Portsmouth, Burnaby Road, Portsmouth, PO1 3FX, UK}

\author[b]{Tian Li,}

\author[a]{Pierre Fleury}

\emailAdd{pierre.fleury@lupm.in2p3.fr}

\abstract{Weak gravitational lensing perturbations have a non-negligible impact on strong lensing observables, and several degeneracies exist between the properties of the main lens, line of sight, and cosmology. In this work, we consider the impact of the line of sight on double-source-plane lenses (DSPLs), a rare class of lens systems in which two sources at different redshifts are lensed by the same foreground galaxy, and which enable competitive constraints on the dark energy equation of state. Generating and sampling statistically representative lines of sight from N-body simulations, we show that line-of-sight perturbations add a $\sim1\%$ uncertainty to measurements of the cosmological scaling factor $\eta$ (a ratio of angular diameter distance ratios), which is subdominant but non-negligible compared to the measurement error. We also show that the line-of-sight shear experienced by images of the two sources can differ significantly in both magnitude and direction. Including a line-of-sight error budget, we measure $w=-1.17^{+0.19}_{-0.21}$ from the Jackpot DSPL in combination with \textit{Planck}. We show that the line of sight is expected to introduce an additional scatter in the constraints possible with a larger sample of DSPLs from \textit{Euclid}, but that this scatter is subdominant compared to other sources of error.}


\date{\today}

\begin{document}

\maketitle
\flushbottom

\section{Introduction}
\label{sec:introduction}
%
Over the last few decades, a rich and varied array of cosmological probes have converged on what is now the standard model of cosmology: a spatially flat universe undergoing an accelerating expansion, dominated by cold dark matter and dark energy in the form of a cosmological constant~\cite{Planck_2020,Brout_2022}. Nonetheless, questions remain to be addressed before this model is universally accepted. Key tensions have emerged between early and late universe measurements of certain parameters, most notably $H_0$ and $\sigma_8$~\cite{Verde_2019, Heymans_2021,Di_Valentino_2021,Hu_2023,Abdalla_2022}. Most probes suffer from astrophysical uncertainties, and these tensions limit the confidence with which other parameters can be constrained. Furthermore, tight constraints on dark energy dynamics have not yet been achieved, and recent results from the DESI collaboration seem to favour a redshift-dependent equation of state over the $\Lambda$CDM paradigm~\cite{DESI_2024,DESI_2024b}. 

Strong gravitational lensing is unique amongst the observational tools available to cosmologists, as lensing observables are sensitive to distances between two non-zero redshifts, thus providing a novel probe of the expansion history of the universe. In rare cases, a foreground galaxy may form images of two background sources at different redshifts, resulting in a so-called double-source-plane lens (DSPL). Even if this foreground galaxy dominates the lensing potential, and thus the background sources are lensed by the same mass, their different angular diameter distances ensure that the angular separation scale of their images will differ. The ratio of their Einstein radii can be measured with high precision, and allows cosmological parameters to be constrained independently of the Hubble constant~\cite{Collett_2012,Sharma_2023}.

The first observation of such a system, The `Jackpot' lens J0946+1006, was identified by~\cite{Gavazzi_2008} with the Sloan Digital Sky Survey~\cite{Bolton_2008} and follow ups with the Hubble Space Telescope. Later, a third, more distant multiply-imaged source was observed as part of the same system~\cite{Collett_2020,Smith_2022}, making J0946+1006 also the first triple source system. This system has been used to place constraints on $\Omega\e{m}$ and the dark energy equation of state parameter~$w$. Most recently, re-analysis of the known galaxy-scale strong lensing system J1721+8842 revealed that the six observed quasar images arose from the same background source, the first confirmation of an Einstein zig-zag lens~\cite{Dux_2024}. 

While only five other DSPL systems have been spectroscopically confirmed~\cite{Tu_2009,Tanaka_2016,Schuldt_2019,Lemon_2023,Bolamperti_2023}, upcoming surveys are expected to increase this number by at least two orders of magnitude~\cite{Sharma_2023}. The Legacy Survey of Space and Time (LSST) \cite{LSST_2021} and the \textit{Euclid} \cite{Euclid_2022} telescope are both forecast to substantially increase the strong lens population \cite{Collett_2015}. Euclid is particularly powerful for DSPL discovery, since it will survey roughly a third of the entire sky with a 0.1 arcsecond resolution, unprecedented for such a wide survey. Identifying a DSPL likely requires the Einstein radii of each source to differ by more than the angular resolution, so Euclid's excellent resolution offers a major advantage for DSPL identification. Euclid is forecast to detect $\sim$ 1700 such systems \cite{Li25}, and 4 have been serendipitously discovered in the initial Q1 area \cite{Li25} (which is a representative patch of 64 deg$^2$ of the final $\sim$140000 deg$^2$ Euclid wide survey \cite{Aussel_2025}) . Given the high number of systems, the extensive sky coverage and the possibility of precise redshift measurements of lens and source light, \textit{Euclid} holds massive promise for DSPL cosmology \cite{Sharma_2023}.

In recent years, increasing attention has been paid to the impact on strong lensing observables of inhomogeneities along the line of sight, and it is now understood that, for precision applications of strong lensing, these effects cannot be ignored~\cite{Wong_2010,Wong_2011,Collett_2013,Greene_2013,Fleury_2019a}. Gravitational shear effects arising from matter external to the main lens~\cite{Gunn_1967,Kochanek_1988,Jaroszynski_1990,Jaroszynski_1991,Seljak_1994,Bar-Kana_1996} can be as strong as the shear produced by the main lens itself~\cite{Bar-Kana_1996}, and a few percent of the total lensing distortion~\cite{Schneider_2019}. Consequently, azimuthal weak lensing effects must be included in models of strong lenses, and their omission can introduce biases~\cite{Jaroszynski_2014,McCully_2017,Johnson_2024} or prevent successful modelling~\cite{Jaroszynski_2014}. The most common approach is to include a tidal ``external shear'' term, meant to absorb these effects, as well as any shear not reproduced by the mass model of the main lens~\cite{Kovner_1987,Schneider_1992,Seitz_1994,Bar-Kana_1996}.

The most challenging consequence of matter external to the main lens model is in the so-called mass sheet degeneracy (MSD). Lensing observables, other than time delays, are unaffected by the addition of an infinite sheet of mass to the lensing potential, up to a rescaling of the lens mass and source plane ~\cite{Falco_1985,Gorenstein_1988,Saha_2000,Schneider_2013}. This degeneracy prevents the precise determination of the mass contained both within the main lens plane and along the line of sight, and is currently a large source of uncertainty in $H_0$ measurements from time-delay cosmography~\cite{Bonvin_2016,Millon_2020,Kumar_2015,Tewes_2013,Chen_2019,Suyu_2013,Yildirim_2021,Treu_2023}. 

The relevance of the line of sight extends to the domain of multi-plane lensing, and extensive work has been done to incorporate these effects into the multi-plane lensing formalism~\cite{McCully_2014,McCully_2017,Fleury_2021a,Fleury_2021b,Schneider_2019}. While multi-source plane lensing partially breaks the mass-sheet degeneracy~\cite{Bradac_2004}, some residual uncertainty is unavoidable~\cite{Liesenborgs_2008,Schneider_2014,Teodori_2022}. 

A key consequence of the MSD is that it is generally impossible to separate contributions to the Einstein radius by the main lens itself from those by matter along the line of sight. Due to the additional distance to the background source in a DSPL system, this source is generally subject to different integrated line-of-sight effects than the lower redshift source, and thus the scaling parameter measured in DSPLs is subject to an inherent uncertainty which does not cancel out, and which cannot be mitigated by lensing observables alone. Furthermore, this additional line of sight means that the external shear acting on light from the two sources will not, in general, be equal. 

In light of the greater than hundredfold increase in the number of identified DSPLs anticipated with current and near future surveys, the combination of cosmological DSPL measurements will be much more constraining than today. With these improvements comes a need to carefully understand systematics and sources of error hidden in this data. The primary motivation of this work is therefore to carefully explore and quantify the impact of the line of sight, with the goal of ensuring a realistic error budget in DSPL observables, and understanding the consequences of this additional error on the cosmological constraints possible with current and future DSPL observations. 

This paper is organised as follows: in \cref{sec:theory}, we review the theory of double source plane lensing and its use in constraining cosmology, and show how line-of-sight inhomogeneities affect DSPL observables. In \cref{sec:los_simulations}, we present our methods for simulating realistic tidal line of sight effects for DSPLs with arbitrary redshifts, and in \cref{sec:DSPL_forecasts}, we present the sample of DSPLs expected to be observed with \textit{Euclid}. In \cref{sec:lensing_observables}, we quantify the impact of the line of sight on the cosmological scaling parameter and the line of sight shear observed from DSPL systems. In \cref{sec:cosmological_consequences}, we present flat-$w$CDM cosmological constraints from the Jackpot lens with an error budget updated to include a contribution from the line of sight, and then explore the impact of the line of sight on flat-$w$CDM and $w_0w_a$CDM cosmological constraints from the forecasted \textit{Euclid} DSPLs. We summarise our results in \cref{sec:conclusion}.

\section{Theoretical background}
\label{sec:theory}

\subsection{Double source plane lensing}
\label{sec:lensing_basics}
The lens equation relates the apparent angular position $\btheta$ of a light ray to the position $\bbeta$ at which it would be observed in the absence of lensing. If the path of the ray between observer o and source s is affected only a single lens plane d and the underlying geometry of the universe, then the difference between these quantities is governed by the deflection angle~$\uvect{\alpha}$ of d, and the lens equation can be written as
\begin{equation}
    \bbeta = \btheta - \frac{D\e{ds}}{D\e{s}}\uvect{\alpha}(\btheta)= \btheta-\balpha(\btheta),
    \label{eq:lens_equation}
\end{equation}
where $D_{ij}$ will refer throughout to the angular diameter distance from planes $i$ to $j$, and we have introduced the displacement angle $\balpha(\btheta)$. It is often convenient to write $\balpha(\btheta)$ in terms of a scalar potential $\psi(\btheta)$, such that
\begin{equation}
    \balpha(\btheta) = \frac{\rd \psi(\btheta)}{\rd\btheta}.
\end{equation}
\subsubsection{The cosmological scaling factor}
Suppose, instead of a single source plane s$_1$ at $z\e{s_1}$, we observe an additional second source s$_2$ at $z\e{s_2}>z\e{s_1}$. For the time being, we will suppose that the light from this second source is lensed only by the foreground lens galaxy, and neglect any additional deflection as its light passes the source at $z\e{s_1}$. This assumption is discussed further in \cref{sec:compound_lensing}. In this case, $\uvect{\alpha}$ is unchanged for the images corresponding to both sources, but the prefactor ratio of angular diameter distances is affected. The ratio of these prefactors is given by the cosmological scaling factor $\eta$,\footnote{Instead of $\eta$, $\beta=\eta^{-1}$ is commonly used in the literature, but the use of $\eta$ is more convenient for lens modelling \citep{Ballard_2024}.}
\begin{equation}
    \eta \equiv \frac{D\e{s_1}D\e{ds_2}}{D\e{ds_1}D\e{s_2}}.
    \label{eq:eta_definition}
\end{equation}
For isothermal primary lens models and no mass on the first source, this quantity can be inferred directly from the ratio of Einstein radii corresponding to s$_2$ and s$_1$, and thus it is readily observed from lens images \citep{Collett_2012}. In real lenses, the situation is not as simple: a lens model must be constructed that fits the normalization and density profiles of both the lens and first source. The $\eta$ parameter enters into the model in rescaling all of the physical deflection angles of the primary lens into the ratio of reduced deflection angles for the two source planes \citep{Collett_2014}.

\subsubsection{Cosmology with DSPLs}
The angular diameter distances appearing in \cref{eq:eta_definition} are given by
\begin{equation}
    D_{ij} = \frac{c}{(1+z_j)}\frac{1}{H_0\sqrt{|\Omega_k|}}\mathrm{sinh}\left(\sqrt{|\Omega_k|}\int^{z_j}_{z_i}\frac{\rd z}{E(z)}\right),
    \label{eq:angular_diameter_distance}
\end{equation}
%
For a $w$CDM cosmology (with $w$ constant but not necessarily equal to $-1$), the normalised Hubble parameter $E(z) \equiv H(z)/H_0$ is given by
\begin{equation}
    E^{w\mathrm{CDM}}(z) = \sqrt{\Omega\e{m}(1+z)^3+\Omega_k(1+z)^2+\Omega\e{de}(1+z)^{3(1+w)}}.
\end{equation}
From \cref{eq:eta_definition}, we see that the dependence of $\eta$ on $H_0$ cancels out, and it therefore depends only on $w$, $\Omega\e{m}$, $\Omega_k$ and the redshifts of the lens and sources. Furthermore, because the Einstein radii from which $\eta$ is constrained are generally measurable to high precision, DSPLs can offer competitive cosmological constraints, and have been used as effectively in combination with other probes to constrain the dark energy equation of state parameter $w$~\cite{Collett_2014,Smith_2022}.

\subsection{Line-of-sight effects}
\label{sec:los_effects}
Real lenses do not exist in isolation, but are instead affected by the distribution of matter along the line of sight between the observer, lens and source. In the presence of purely tidal line-of-sight perturbations and a single dominant lens, the lens equation takes the form
\begin{equation}
    \bbeta=\bA_{\os}\btheta-\bA_{\ds}\frac{\rd\psi(\bA_{\od}\btheta)}{\rd\btheta},
    \label{eq:DL_lens_tidal_beta'}
\end{equation}
where $\psi$ here refers to the lensing potential of the main lens only~\cite{Kovner_1987,Bar-Kana_1996,Schneider_1997}. The amplification matrices appearing in the above expression describe the distortions to an infinitesimal beam of light due to foreground perturbers between the observer and lens ($\bA_{\od}$), background perturbers between lens and source ($\bA_{\ds}$) and perturbers between observer and source ($\bA_{\os}$). A full derivation of this expression can be found in~\cite{Fleury_2021a}.
The amplification matrices appearing in \cref{eq:DL_lens_tidal_beta'} are typically parameterised as 
\begin{equation}
    \bA_{ij} = \begin{pmatrix} 1-\kappa_{ij}-\gamma_1^{ij}& -\gamma_2^{ij} \\ -\gamma_2^{ij} & 1-\kappa_{ij} + \gamma_1^{ij} \end{pmatrix},
    \label{eq:a_matrix}
\end{equation}
where the convergence $\kappa_{ij}$ is the surface mass density in units of the respective critical density, which symmetrically rescales an image, and the shear $\gamma_{ij}=\gamma_1^{ij}+\ri\gamma_2^{ij}$ results in non-symmetric distortions. In the following, we will also refer to the reduced shear $g_{ij} = (1-\kappa_{ij})^{-1}\gamma_{ij}$.
\subsection{The line of sight and double source plane observables}
\label{sec:double_plane_los}
\subsubsection{The cosmological scaling factor}
The expression for $D_{ij}$ in \cref{eq:angular_diameter_distance} is exactly true for a universe which is homogeneous and isotropic. However, in the presence of matter inhomogeneities, the resulting weak lensing convergence $\kappa_{ij}$ acts to uniformly change the angular size of an object at a given redshift, as over(under)dense lines of sight increase (decrease) the focusing of a light beam. Its effect on a beam can therefore be thought of as a transformation of angular diameter distances, $D_{ij} \rightarrow (1-\kappa_{ij})D_{ij}$, and thus, irrespective of the main lens model, convergence along the line of sight will bias the $\eta$ parameter by a factor of
\begin{align}
    \frac{\eta\h{inferred}}{\eta} &= \frac{(1-\kappa\e{s_1})(1-\kappa\e{ds_2})}{(1-\kappa\e{ds_1})(1-\kappa\e{s_2})}.
    \label{eq:eta_bias}
\end{align}
This fact remains true with or without the lensing effects of the galaxy at s$_1$, provided that the quantity being measured is indeed~$\eta$ as given by \cref{eq:eta_definition}. However, mass in s$_1$ has the potential to complicate the interpretation of the observed Einstein radii, and hence the extraction of~$\eta$. The details of this discussion are presented in \cref{sec:compound_lensing}, but the key point is that, unsurprisingly, the mass in s$_1$ can be safely ignored only if the Einstein radius of s$_1$ for the source s$_2$ is negligible compared to that of d for the same source. For isothermal lenses, this translates to the requirement that 
\begin{equation}
    \frac{D\e{s_1s_2}}{D\e{ds_2}} \ll \frac{\sigma_{v,\mathrm{d}}^2}{\sigma_{v,\mathrm{s}_1}^2},
\end{equation}
where the $\sigma_v^2$ are the velocity dispersions of d and s$_1$. Clearly, this inequality need not be satisfied in general.

If the lensing effect of s$_1$ is sufficiently strong, the system can produce three Einstein rings, and the average of the inner and outermost radii give us the value needed to obtain~$\eta$~\cite{Collett_2016}. Even if the mass in s$_1$ is insufficient to produce an additional ring or additional images of s$_2$, it may nonetheless be constrained through its effect on lensing observables, as, for example, in the case of the Jackpot lens~\cite{Collett_2014}. Thus, $\eta$ is readily measurable if the mass in s$_1$ is sufficiently large or small, while in the intermediate cases its influence may be less separable. Nonetheless, the central point remains -- if the measured quantity is indeed~$\eta$ as defined in \cref{eq:eta_definition}, the effects of weak lensing on angular diameter distances ensure that this quantity will be biased according to \cref{eq:eta_bias}.
\subsubsection{The line-of-sight shear}
\label{sec:shear_effects}
We can also consider the consequences of the line of sight when modelling the shear of the lens system. When modelling a lens, a so-called `external shear' parameter is often included to improve the image reconstruction. This term is something of a misnomer, and this `residual shear'~\cite{Tan_2024} captures the effects of both matter along the line of sight, and mass within the main lens plane which is not fully captured by the model~\cite{Etherington_2023}. Furthermore, as a consequence of the mass-sheet degeneracy, the line-of-sight convergence of the system is typically implicitly set to 0, which means that the shear being measured is in fact the reduced shear. 

When modelling the residual shear on a double source plane lens system, it is sometimes implicitly assumed that the shear exists entirely in the main lens plane. Under this assumption, the shear acting on light from s$_2$,  $g\e{LOS,2}$, is simply a displacement term proportional to $D\e{ds_2}/D\e{s_2}$, and thus related to the shear on s$_1$, $  g\e{LOS,1}$ by a factor of $\eta$, 
\begin{equation}
    g\e{LOS,2} = \eta \, g\e{LOS,1}, 
\end{equation}
such as in~\cite{Collett_2014,Schuldt_2019,Bolamperti_2023,Ballard_2024,Enzi_2024}. In real lensing systems, however, there is no reason that this should be the case, as the line-of-sight contribution to the residual shear may change substantially as a function of source redshift. We quantify these effects in \cref{sec:lensing_observables}.

\section{Simulating lines of sight}
\label{sec:los_simulations}
%
\begin{figure}
    \centering
    \includegraphics[width=1\textwidth]{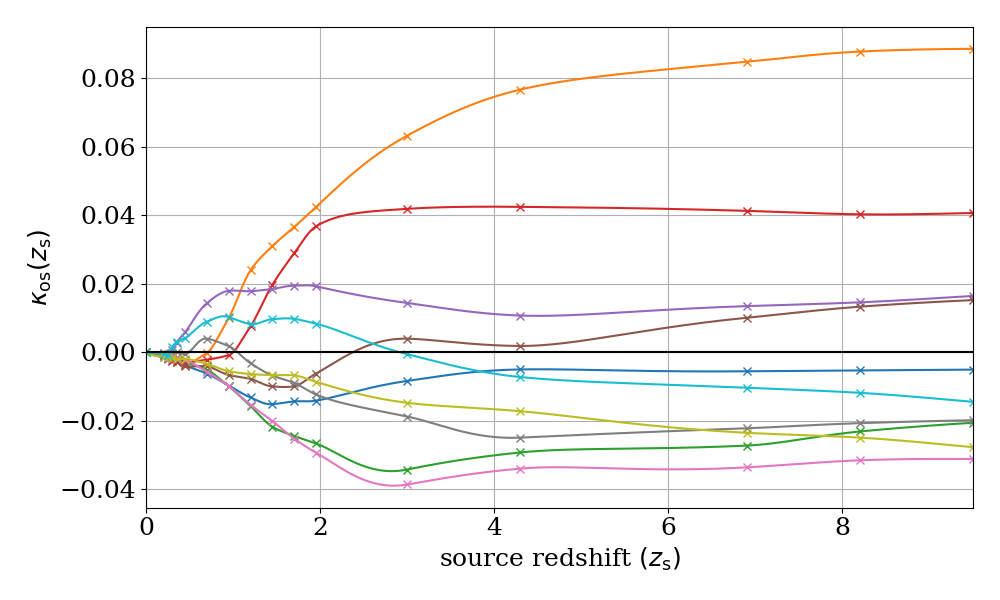}
    \caption{$\kappa\e{os}(z\e{s})$, the convergence between the observer at $z=0$ and a source at $z\e{s}$, for a random sample of 10 lines of sight. The datapoints taken from the RalGalGroup simulations are plotted as crosses, and the interpolated convergence between them as solid lines.}
    \label{fig:kappa(z)}
\end{figure}
In order to quantify the effect of the line of sight on double source plane lens systems, we must estimate $\kappa_{ij}$ and $\gamma_{ij}$ between redshifts $z_i$ and $z_j$ for arbitrary but representative lines of sight. From the results of RayGalGroupSims~\cite{Breton:2018wzk}, which carries out fully relativistic ray tracing in a dark-matter simulation based on the $N$-body code RAMSES~\cite{Teyssier_2002, Guillet_2011} using the \textsc{Magrathea} library~\cite{reverdy2014propagation, Breton}, we have access to $\kappa\e{\os}$ and $\gamma\e{os}$ at different values of the source redshift. The simulation uses a comoving length of $2625h^{-1}~\text{Mpc}$ and a particle mass of $1.88 \times 10^{10}h^{-1} M_\odot$ within a WMAP-7 cosmology~\cite{komatsu2011seven}.\footnote{We do not expect the magnitude of the effect to change significantly if redetermined in an updated cosmological model, and these results are sufficient to quantify the relevance of the line of sight.} The resulting HEALPix convergence, shear and magnification maps are publicly available.\footnote{\href{https://cosmo.obspm.fr/public-datasets/raygalgroupsims-relativistic-halo-catalogs}{\tt https://cosmo.obspm.fr/public-datasets/raygalgroupsims-relativistic-halo-catalogs}} By determining the corresponding comoving distances $\chi\e{s}$ at these redshifts, we can interpolate these values to get $\kappa\e{os}$ and $\gamma\e{os}$ as a continuous function of $\chi\e{s}$, from which we can determine any convergence or shear term as seen from the observer at $z=0$ (ie. $\kappa\e{os},\gamma\e{os}$, $\kappa\e{od}$, and $\gamma\e{od}$ in the case of a single dominant lens plane) for a randomly sampled line of sight along which we perform this interpolation. \cref{fig:kappa(z)} shows the convergence between an observer at $z=0$ and a source located at $z\e{s}$ for a random sample of lines of sight.

This, however, is insufficient, as lensing observables are also affected by terms such as $\kappa\e{ds}$ and $\gamma\e{ds}$. What we therefore need is a function to determine $\kappa_{ij}$ and $\gamma_{ij}$, the convergence and shear between a pair of planes $i$ and $j$ at arbitrary redshifts, as seen from plane $i$. 

References ~\cite{Fleury_2019a,Fleury_2021b} give expressions for $\kappa_{ij}$ and $\gamma_{ij}$ in terms of the density contrast $\delta$. If we invert these to express $\delta$ as a function of $\kappa\e{os}$, we can obtain
\begin{align}
    \kappa_{ij} &= \int^{\chi_j}_{\chi_i}\frac{\rd\chi}{\chi}\;W_{ij}\frac{\rd^2}{\rd\chi^2}\left[\chi\kappa_\os(\chi)\right], \\
    \gamma_{ij} &= \int^{\chi_j}_{\chi_i}\frac{\rd\chi}{\chi}\;W_{ij}\frac{\rd^2}{\rd\chi^2}\left[\chi\gamma_\os(\chi)\right],
\end{align}
where the weighting function $W_{ij}$ is given by
\begin{equation}
    W_{ij} \equiv \frac{(\chi_j-\chi)(\chi-\chi_i)}{\chi_j-\chi_i}.
\end{equation}
Thus, given functions $\kappa\e{os}(\chi)$, $\gamma\e{os}(\chi)$ from simulations, we can obtain the corresponding value of $\kappa\e{os}$ and $\gamma_{ij}$ between comoving distances $\chi_i$ and $\chi_j$. 

\section{Forecasted double source plane lens populations}
\label{sec:DSPL_forecasts}

\textit{Euclid} is expected to contain a sample of $\sim1700$ galaxy-scale DSPL systems that could plausibly be used to constrain cosmological parameters \citep{Li25}. This forecast was derived from the {\sc LensPop} package \citep{Collett_2015} modified to include multiple background sources and neglecting mass in the first source. This sample only includes systems where both sources have one or more arcs of length $\ang{;;0.3}$, which  ensures a reasonable possibility that the density slope of the lens can be recovered from \textit{Euclid} imaging alone. In this work, we will assume that Euclid is able to find all of these forecasted lenses, with redshifts and Einstein radii following \citep{Li25}. As in \citep{Li25}, we will assume that each DSPL enables a measurement of $\eta$ with an uncertainty given by the quadrature sum of 0.01 arcseconds over the Einstein radius of the first source, the same for the second source and 0.01. The first two terms represent reasonable uncertainties on the Einstein radius determined from lens modelling, whilst the last term sets a floor on the uncertainty given that the density profile slope of the primary lens is not perfectly known.

\section{Quantified effects on lensing observables}
\label{sec:lensing_observables}

\subsection{A single system - the Jackpot lens}
\label{sec:jackpot_observables}
\begin{figure}
    \centering
    \includegraphics[width=1\textwidth]{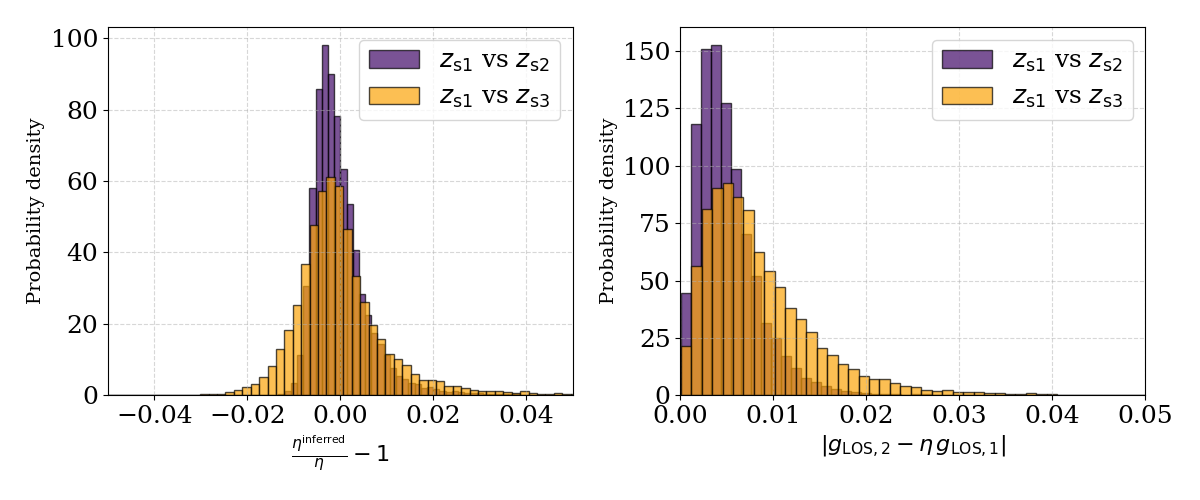}
    \caption{On the left, the probability density function for $\eta^\mathrm{inferred}/\eta-1$, and on the right, the probability density function for $|g\e{LOS,2}-\eta \, g\e{LOS,1}|$, for the Jackpot lens. In blue, we consider values corresponding to the first and second source planes ($z\e{s_1} = 0.609$, $z\e{s_2} = 2.035$), and in orange, we consider values corresponding to the first and third source planes ($z\e{s_1} = 0.609$, $z\e{s_2} = 5.975$). The standard deviation of the bias on $\eta$ is 0.67\% for the inner pair of rings, and 1.3\% for the outer pair.}
    \label{fig:histograms_jackpot}
\end{figure}

The `Jackpot' lens, J0946+1006, remains the most famous example of multi-source plane lensing. Discovered by~\cite{Gavazzi_2008} as part of the SLACS survey~\cite{Bolton_2006,Bolton_2008}, it was initially identified as a double source plane lens, with spectroscopically-confirmed redshifts of $z\e{d}=0.222$, $z\e{s_1}=0.609$ and $z\e{s_2}=2.035$~\cite{Gavazzi_2008,Smith_2022}. More recently, a third multiply-imaged source has been confirmed at $z\e{s_3}=5.975$~\cite{Collett_2020}. 

Using the redshifts of the Jackpot lens and its sources, and the methods described in \cref{sec:los_simulations}, we generate 20 000 lines of sight, and extract the relevant convergence and reduced shear terms. For each of these lines of sight, we calculate the bias on $\eta$ as given by \cref{eq:eta_bias}, as well as the magnitude of the difference between the actual ``external shear'' of the outer ring (corresponding either to s$_2$ or s$_3$) and the rescaling of the shear determined from the innermost ring. The results of this are plotted in \cref{fig:histograms_jackpot}. 

From the figure on the left, we see that the line of sight typically introduces a bias on $\eta$ on the order of a percent, with a standard deviation of $\eta\h{inferred}/\eta$ of 0.67\% for the inner pair of rings, and 1.3\% for the outer. While the median of the bias is $\approx 0$, the histograms are not perfectly symmetric, and large overestimations of $\eta$ are more frequent than large underestimations. Unsurprisingly, we see that the typical size of the bias increases when considering the value of $\eta$ as calculated between s$_1$ and s$_3$, as the larger redshift differences between these two sources mean the line of sight becomes, on average, more significant (see \cref{fig:kappa(z)}). For comparison, the measurement error in $\eta$ presented in~\cite{Collett_2014} is 1.1\%. That statistical error thus remains the dominant source of uncertainty, but the effect of the line of sight is not completely negligible. While our lines of sight are chosen without accounting for selection effects, some simple tests suggest that such effects may add a small additional bias to measurements of $\eta$, but that this bias is unlikely to be much larger than the scatter seen in \cref{fig:histograms_jackpot}. This is discussed in more detail in \cref{sec:selection_effects}.

From the figure on the right, we see that the difference between the line-of-sight shear acting on the inner and outer rings is typically slightly smaller than a percent, once again growing larger when comparing with the more distant source. While these values are small in absolute terms, this corresponds to a median difference of $\sim$140\% of $\eta \, g\e{LOS,1}$ in the case of s$_2$, and $\sim$230\% in the case of s$_3$. It is therefore clear that $\eta \, g\e{LOS,1}$ is a poor substitute when modelling $g\e{LOS,2}$ and $g\e{LOS,3}$. Compound lens models should allow for an additional external shear term on each lens plane.

\subsection{A sample of mock \textit{Euclid} DSPLs}
\begin{figure}
    \centering
    \includegraphics[width=1\textwidth]{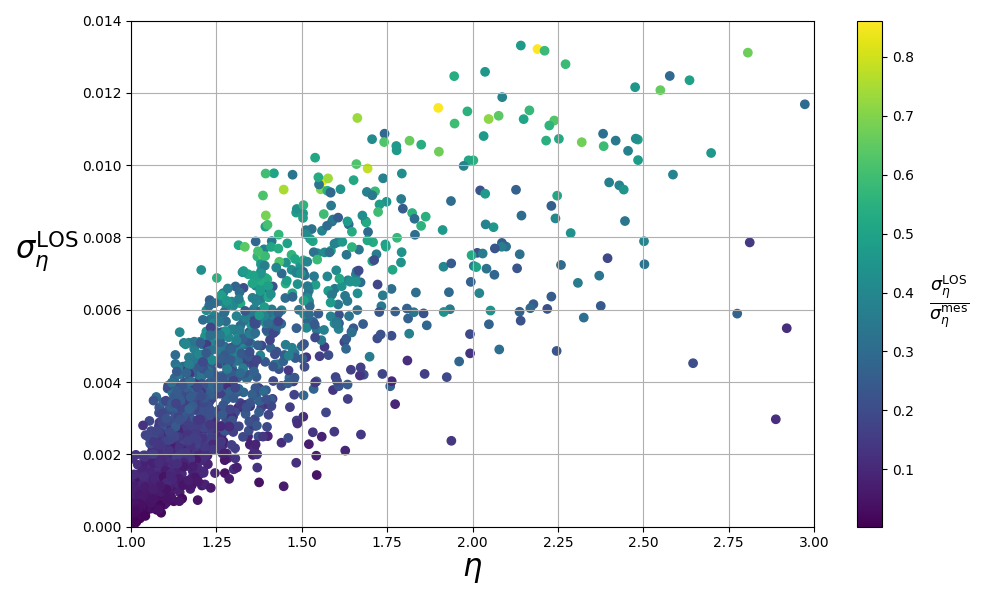} 
    \caption{Expected fractional error on $\eta$ arising from the line of sight as a function of $\eta$, for a sample of $\sim 1700$ forecasted \textit{Euclid} DSPLs. The colour of the points shows this error as a fraction of the error on $\eta$ from measurement uncertainties.}
    \label{fig:scatterplot_etas_Euclid}
\end{figure}
To better understand the significance of the line of sight as a function of the redshifts of the lens and sources, we repeat the procedure described in \cref{sec:jackpot_observables} for a sample of 1729 systems forecasted to be observable with \textit{Euclid} (\cref{sec:DSPL_forecasts}). For each of these systems, we simulate 1000 lines of sight and calculate the biases on $\eta$ and $g\e{LOS}$ which would arise from ignoring these line-of-sight effects.
\begin{figure}
    \centering
    \includegraphics[width=1\textwidth]{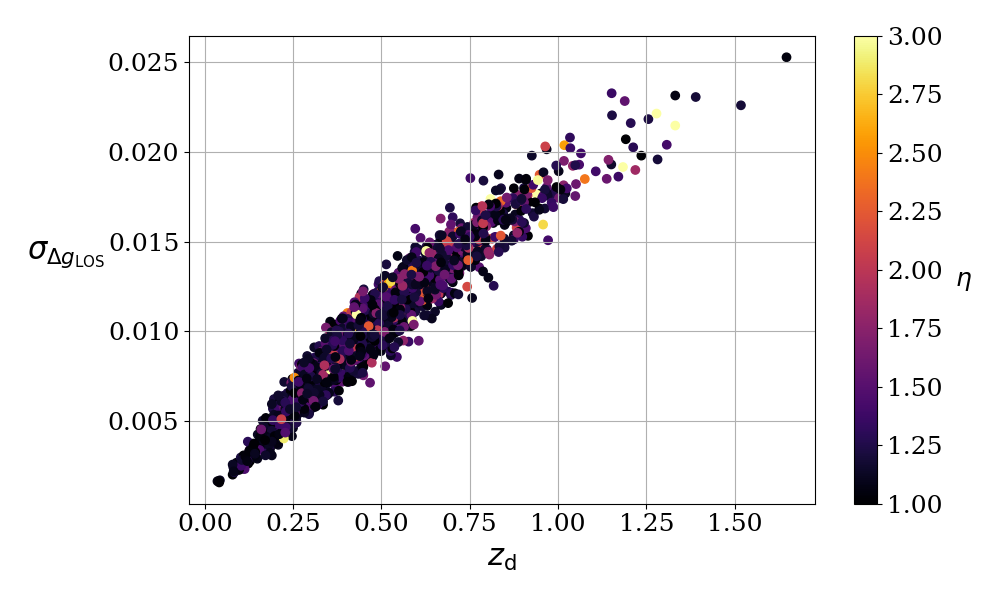} 
    \caption{Expected differences arising from the line of sight between $g\e{LOS,2}$ (the true ``external shear'' acting on the images of s$_2$) and $\eta \, g\e{LOS,1}$ for a sample of $\sim1700$ forecasted \textit{Euclid} lenses, as a function of the deflector redshift $z\e{d}$, with $\eta$ shown as the colours of the points.}
    \label{fig:scatterplot_gammas_Euclid}
\end{figure}

The expected biases on $\eta$ are shown in \cref{fig:scatterplot_etas_Euclid}. Here, rather than considering the full PDF for each lens, we summarise the results in the parameter $\sigma\h{LOS}_{\eta}$, which is half the difference between the 84$\h{th}$ and 16$\h{th}$ percentiles for $\eta\h{inferred}/\eta$ across the 1000 lines of sight per lens system.\footnote{Considering the percentile difference rather than the standard deviation prevents $\sigma\h{LOS}_\eta$ being dominated by a small subset of outliers.} We use the colours of the datapoint to illustrate this parameter as a fraction of $\sigma\h{mes}_\eta$, which is the forecasted fractional measurement error on $\eta$ coming from errors unrelated to the line of sight (such as resolving the Einstein radius and disentangling the effects of the mass in s$_1$). From the figure, we see that the $\sigma\h{LOS}_{\eta}$ grows as $\eta$ grows, and, for systems with higher $\eta$ values, $\sigma\h{LOS}_{\eta}$ can often be comparable with the measurement error on $\eta$. This point is particularly relevant given that the higher $\eta$ systems are the more powerful when constraining cosmology, and thus of greater scientific interest~\cite{Collett_2012}.\footnote{We expect $\sigma\h{LOS}_\eta$ to be a precise function of $z\e{d}$, $z\e{s_1}$ and $z\e{s_2}$, and the dispersion in \cref{fig:scatterplot_etas_Euclid} arises because $\eta$ does not fully capture the variation in these quantities.}

In \cref{fig:scatterplot_gammas_Euclid}, we have summarised the differences between $g\e{LOS,2}$ and $\eta \, g\e{LOS,1}$, calculated for different lines of sight, via the quantity $\sigma_{\Delta g_\mathrm{LOS}}$, which is half the difference between the 84$\h{th}$ and 16$\h{th}$ percentiles of the magnitude of that difference across the 1000 lines of sight considered per lens system. There is a strong correlation between this quantity and $z_\mathrm{d}$, but there is only a weak correlation with $\eta$. This contrasts with the $\sigma_{\Delta g_\mathrm{LOS}}$ parameter, which correlated most strongly with $\eta$. Thus, the importance of the additional line-of-sight shear acting on s$_2$ is determined mainly by the overall scale of the DSPL, rather than by the relative positions of s$_1$, s$_2$, and $d$. Nonetheless, it is clear that the magnitude of this additional shear, while significant relative to the shear on s$_1$, is small relative to the lensing effect of the main lens.  

\section{Impact on cosmology}
\label{sec:cosmological_consequences}
%
\begin{figure}
    \centering
    \includegraphics[width=1\textwidth]{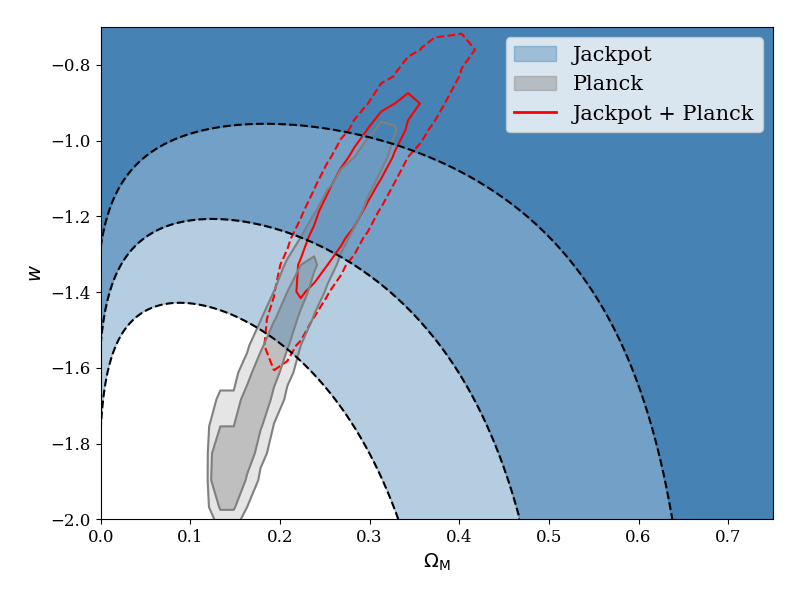} 
    \caption{Constraints on $w$ and $\Omega\e{m}$ in a flat-$w$CDM cosmology. Shown in blue are the constraints from the Jackpot lens alone, where the uncertainty from the line of sight added in quadrature to the measurement error reported in~\cite{Smith_2022}. In gray are the constraints from \textit{Planck}~\cite{Planck_2020}, and in red are the combined constraints from the Jackpot lens and \textit{Planck}. The combined constraints give a value of $w=-1.17^{+0.19}_{-0.21}$.}
    \label{fig:combined_constraints}
\end{figure}
Measurements of the quantity $\eta$ from DSPLs allow us to place constraints on cosmological quantities such as $\Omega\e{m}$, the matter density parameter, and $w(z)$, the dark energy equation of state~\cite{Collett_2012,Sharma_2022}. This has motivated the work of~\cite{Collett_2014,Smith_2022}, and the anticipated samples of DSPLs from LSST and \textit{Euclid} promise to offer new and competitive constraints~\cite{Sharma_2023}. In this section, we aim to explore and quantify the effect of the line of sight on the cosmological constraints possible from DSPLs.

\subsection{Updated constraints from the Jackpot lens}

Using $\eta = 1.405^{+0.014}_{-0.016}$, $z\e{d}=0.222$, $z\e{s_1}=0.609$ and $z\e{s_2}=2.41^{+0.04}_{-0.21}$,~\cite{Collett_2014} obtained constraints on $\Omega\e{m}$ and $w$ for a flat $w$CDM cosmology, which were updated in~\cite{Smith_2022} with the new spectroscopic redshift measurement for the background source, $z\e{s_2}=2.035$. These constraints, however, did not take into account the additional uncertainty on $\eta$ arising from the line of sight. Following the results presented in \cref{sec:jackpot_observables}, we add in quadrature a 0.67\% error budget for the line of sight to the measurement uncertainty presented in~\cite{Smith_2022}, and present the updated constraints in Figure \ref{fig:combined_constraints}. Also shown are the constraints from \textit{Planck}~\cite{Planck_2020}, and the combined constraints from \textit{Planck} and the Jackpot lens with the line-of-sight error included, giving a dark energy equation of state parameter of $w=-1.17^{+0.19}_{-0.21}$. Using the same data but not including the line-of-sight error, the constraint is instead $w=-1.12^{+0.17}_{-0.18}$, meaning that the line of sight adds $\sim 15\%$ to the uncertainty of the final result.
\subsection{Constraints from a large sample of DSPLs}
\begin{figure}
    \centering
    \includegraphics[width=1\textwidth]{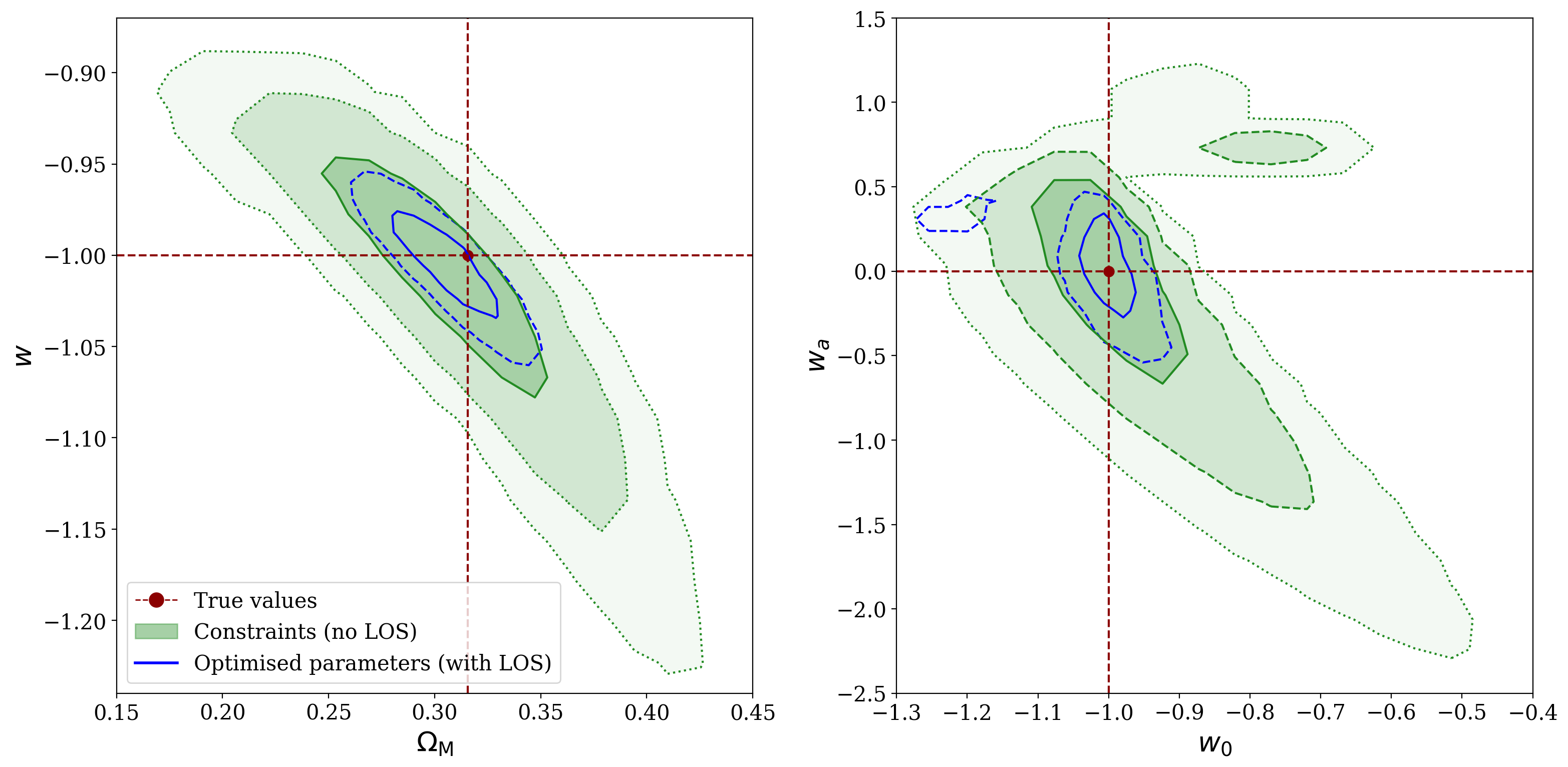} 
    \caption{The scatter introduced by the line of sight compared to the uncertainty in cosmological parameter constraints for a $w$CDM  (left) or $w_0w_a$CDM (right) cosmology. In green, we show the 1, 2, and 3-$\sigma$ constraints possible from the forecasted set of 1729 \textit{Euclid} DSPLs and the anticipated measurement uncertainties in $\eta$, with no budget for the line of sight error. In blue, we show the contours for the best guess of these parameters, for 50 000 variations on the lines of sight of each of the DSPLs used in obtaining the constraints. The ``true'' parameter values are plotted in red.}
    \label{fig:Euclid_cosmology_constraints}
\end{figure}
We now quantify the impact of these line-of-sight uncertainties on the cosmological parameter constraints possible from a \textit{Euclid}-like DSPL sample. If the line of sight is not accounted for in the error budget for $\eta$, but nonetheless present in the data, the main effect is to shift the posterior samples within the parameter space, without significantly affecting their shape or width. If the effect is significantly large, the posterior may be shifted such that the true value no longer lies within 1-$\sigma$ of the measurement. However, given the large number of DSPLs, we would also expect these effects to be mitigated, given that the line of sight bias is not systematic.

To investigate this, we firstly consider the constraining power of the forecasted sample of \textit{Euclid} systems and the precision with which their $\eta$ values are expected to be measured. This is shown in green in \cref{fig:Euclid_cosmology_constraints} for a flat $w$CDM cosmology (left), as well as for the CPL parameterisation of a redshift-dependent dark energy equation of state,
\begin{equation}
    w(z) = w_0 + w_a\frac{z}{1+z},
\end{equation}
shown on the right~\cite{CHEVALLIER_2001,Linder_2003}. We then resample these posteriors, but this time generating lines of sight for each of the DSPLs, and updating the $\eta$ values according to \cref{eq:eta_bias}. We extract and save the new optimised cosmological parameters after this resampling. We repeat this process 50 000 times, and plot the distribution of these optimised parameters in blue in \cref{fig:Euclid_cosmology_constraints}.

As expected from \cref{fig:scatterplot_etas_Euclid}, the scatter arising from the line of sight is usually subdominant compared to the anticipated measurement uncertainties on the $\eta$ values. In $\sim 95\%$ of line-of-sight realisations, the centroid of the distribution remains within the 1-$\sigma$ errorbars of the idealised constraints for a flat-$w$CDM cosmology. This is also true in the majority of realisations for a flat-$w_0w_a$CDM cosmology, although, as shown by the small 2-$\sigma$ region in the lower part of the figure on the right in \cref{fig:Euclid_cosmology_constraints}, line-of-sight inhomogeneities are sometimes enough to skew the cosmological constraints more strongly. If line-of-sight effects are neglected, $\sim 35\%$ of realisations of \textit{Euclid} DSPL lines of sight result in a 2-$\sigma$ tension for $w_0$ and $w_a$ compared to their true values. 

It is beyond the scope of this work to fully account for the selection effects which may bias the distribution of lines of sight. Nonetheless, some simple considerations show that, while the number of outliers may be increased when accounting for the effect of line-of-sight influenced Einstein radii on the probability of DSPLs occurring, this does not seem to introduce a noticable systematic bias in cosmological constraints. This is discussed and illustrated in \cref{sec:selection_effects}.

\section{Discussion \& conclusion}
\label{sec:conclusion}

In this work, we have investigated and quantified the impact of weak lensing perturbations on observables in double source plane lens systems. We have pointed out that a measurement of the cosmological scaling factor $\eta$ from DSPL observables is inseparable from a ratio of line-of-sight convergence terms. By sampling random lines of sight from the RayGalGroup n-body simulation, and interpolating these results to simulate realistic tidal weak lensing effects between arbitrary redshifts, we show that this ratio introduces an additional uncertainty of 0.1 to 1.5\% on the cosmological scaling factor for the expected sample of lenses observable with \textit{Euclid}. This error is typically sub-dominant but non-negligible when compared with the expected measurement uncertainty on $\eta$ for these systems. The fractional error becomes larger and increasingly relevant as $\eta$ increases, meaning that it is most relevant for the most cosmologically useful systems. Nonetheless, it has the expected mean of 0 and is not a systematic bias.

We have also quantified the expected differences between the line-of-sight shears acting on the inner and outer Einstein rings of a DSPL system. This difference is typically larger than the inner line-of-sight shear itself, meaning that it does not make sense to model these shears as a single external shear in the main lens plane. Nonetheless, these terms are still highly subdominant compared to the deflections by the main lens itself. 

Finally, we have explored the impact of these biases on the cosmological constraints possible from DSPLs. For a single measurement of $\eta$, the additional uncertainty arising from the line of sight adds a small additional width to the constraints, and we have presented updated flat $w$CDM constraints from J0946+1006, the `Jackpot' lens. Because this additional error is subdominant and expected to be uncorrelated with the measurement error, its impact on these constraints is small. Combining these constraints with those from Planck~\cite{Planck_2020}, we obtain a value of $w=-1.17^{+0.19}_{-0.21}$ for the dark energy equation of state parameter, which corresponds to a 14\% increase in the reported error when comparing to the case where the line of sight is ignored.

For cosmological constraints from a larger population of DSPLs, the line of sight adds an additional scatter to the posterior distributions for parameters such as $\Omega\e{m}$ and $w$, as well as in $w_0$ and $w_a$ for a redshift-dependent equation of state. This scatter, while subdominant compared to the overall uncertainty in the constraints, may nonetheless lead to a ~2-$\sigma$ bias in $w_0$ and $w_a$ in $\sim 35\%$ of cases.

Our investigation of links between the properties of the line of sight to the occurrence and discovery of DSPLs has been limited to some simple illustrative models, and is mostly presented in \cref{sec:selection_effects}. Work remains to be done to thoroughly quantify selection effects at play in double-source-plane lensing, and the tools necessary for this are beyond the scope of this work. While the lensing of point sources is known to be sensitive to these effects \citep{Collett_2016b, Li2024}, they are understood to play a smaller role for lensed extended sources \citep{Collett_2015}. This is consistent with our findings, that the biases introduced tend to be subdominant when compared to the uncertainty arising from the line of sight.

While our work has focused on line-of-sight lensing, efforts are still needed to constrain the density profiles of lens galaxies, ensuring that no internal mass-sheet transformation has been missed during lens modelling. A mass sheet in the lens plane need not correspond in any way to expectations from the distributions of lines of sight on larger scales, and, like the line-of-sight effects discussed in this work, can lead to precise but inaccurate lens reconstructions and cosmological constraints \cite{Schneider_2014}.

In summary, while line-of-sight inhomogeneities impact DSPL observables, these effects are small when constraining cosmology at the current level of measurement precision expected from surveys such as \textit{Euclid}. 

Our paper teaches two lessons for future cosmography with large samples of double source plane lenses.
\begin{itemize}
\item Firstly, compound lens models should allow for an additional external shear term on each lens plane, though amplitudes greater than $\sim$ 0.02 for the second lens plane would be surprising from a purely line-of-sight perspective.
\item Secondly, interpretation of lens model constraints on $\eta$ should include a small contribution from the line of sight in the error budget, but these errors will only become dominant if the precision with which Einstein radii are measured is improved.  
\end{itemize}
Taking these two actions will ensure that double source plane lens systems become a competitive method for constraining the dark energy equation of state.

\section*{Acknowledgements}

DJ acknowledges support by the First Rand Foundation, South Africa, and the Centre National de la Recherche Scientifique of France.
This work has received funding from the European Research Council (ERC) under the European Union’s Horizon 2020 research and innovation programme (LensEra: grant agreement No. 945536). TC is funded by the Royal Society through a University Research Fellowship. For the purpose of open access, the authors have applied a Creative Commons Attribution (CC BY) license to any Author Accepted Manuscript version arising.
PF acknowledges support from the French \emph{Agence Nationale de la Recherche} through the ELROND project (ANR-23-CE31-0002). We thank the anonymous reviewer for their helpful comments on an earlier version of this work.

\section*{\href{https://www.elsevier.com/authors/policies-and-guidelines/credit-author-statement}{CRedIT} authorship contribution statement}

\noindent \textbf{Daniel Johnson:} Methodology, Formal Analysis, Writing - Original draft. 
\textbf{Thomas Collett:} 
Conceptualization, Methodology, Writing - Review \& Editing, Supervision. 
\textbf{Tian Li:} Software.
\textbf{Pierre Fleury:} Methodology, Supervision.

\appendix

\section{Compound lensing}
\label{sec:compound_lensing}
In expressions such as \cref{eq:eta_bias}, we have neglected the lensing effect of s$_1$ on the light from s$_2$. However, this effect is not necessarily small, and has been an important component of models of observed DSPLs~\cite{Collett_2014,Tanaka_2016,Schuldt_2019,Ballard_2024,Enzi_2024}. For a system of an arbitrary number of lens planes $l$ in the presence of tidal line-of-sight effects, the lens equation takes the form
\begin{equation}
    \bbeta = \btheta  - \sum^{N}_{l=1}\bA^{-1}\e{os}\bA\e{ls}\uvect{\alpha}_l(\bx_l),
\end{equation}
which follows from equation (42) in~\cite{Fleury_2021b} under the assumption that the lenses are comoving with the cosmic flow. $\bx_l$ refers to the physical position at which light passes through the $l$th lens plane. If we substitute $N=2$, keep the notation d and s$_1$ to refer to the foreground and background deflectors, and follow~\cite{Fleury_2021a} in introducing the displacement angle
\begin{equation}
    \balpha_{ilj}(\bx_l)\equiv\frac{D_{lj}}{D_{ij}}\hat{\balpha}_l(\bx_l), \label{eq:displacement_angle}
\end{equation}
we obtain
\begin{multline}
    \bbeta = \btheta  - \bA^{-1}\e{os_2}\bA\e{ds_2}\balpha\e{ods_2}(D\e{d}\bA\e{od}\btheta) \\ - \bA^{-1}\e{os_2}\bA\e{s_1s_2}\balpha\e{os_1s_2}\left[\bA\e{os_1}\btheta     - D\e{d}\bA\e{ds_1}\balpha\e{ods_1}(D\e{d}\bA\e{od}\btheta) \right].
    \label{eq:compound_lensing}
\end{multline}
To gain some insight into the importance of the second lens plane, we will assume that both deflectors are singular isothermal sphere (SIS) lenses. The deflection angle of this lens takes the form
\begin{equation}
    \balpha_{ijk}(\btheta) = \frac{D_{jk}}{D_{ik}}4\pi\sigma_{\upsilon,j}^2 \uvect{\theta},
\end{equation}
where $\sigma^j_\nu$ is the velocity dispersion of lens $j$, and $\uvect{\theta}$ is the unit vector in the direction of $\theta$. The magnitude of the deflection, which we identify as the Einstein radius of the SIS, is therefore simply given by 
\begin{equation}
    \ttheta_{\mathrm{E},ijk} = \frac{D_{jk}}{D_{ik}}4\pi\sigma_{\nu,j}^2 = |\balpha|.
    \label{eq:thetaE_SIS}
\end{equation}
Following the discussion in~\cite{Teodori_2022,Johnson_2024}, we use the notation $\tthetaE$ to distinguish this parameter from the observed Einstein radius $\thetaE$, which is also affected by matter along the line of sight.
Neglecting the line-of-sight shear terms and assuming the centres of both lenses are aligned, the Einstein radii we would observe for s$_1$ and s$_2$ would be
\begin{align}
    \theta\e{E}\h{s_1} &= \frac{1-\kappa\e{ds_1}}{1-\kappa\e{os_1}}\ttheta\e{E,ods_1}, \\
    \theta\e{E\pm}\h{s_2} &= \frac{1-\kappa\e{ds_2}}{1-\kappa\e{os_2}}\ttheta\e{E,ods_2} \pm \frac{1-\kappa\e{s_1s_2}}{1-\kappa\e{os_2}}\ttheta\e{E,os_1s_2}.
\end{align}
When we measure a compound lens system, we typically have two observables, $\theta\e{E}\h{s_2}$ and $\theta\e{E}\h{s_1}$. Depending on the geometry system, we may be able to measure $\theta\e{E,os_1s_2}$ (the observed Einstein radius of s$_1$ relative to the observer at o and the source s) directly~\cite{Collett_2016}. In many cases, however, we will not be able to model the background deflector s$_1$, and we will observe only the Einstein radius $\theta\e{E}\h{s_2}$ with contributions from both d and s$_1$.

If we observe \textbf{three Einstein rings}, we can extract the Einstein radius between observer, lens and s$_2$ as
\begin{equation}
    \theta\e{E,ods_2} = \frac{1}{2}\left(\theta\e{E+}\h{s_2}+\theta\e{E-}\h{s_2}\right).
\end{equation}
For a singular isothermal lens, $\eta$ is then simply the ratio of $\theta\e{E,ods_2}$ to $\theta\e{E}\h{s_1}$, and the bias given in \cref{eq:eta_bias} is unchanged.

If, however, we observe \textbf{only two Einstein rings}, i.e. only $\theta\e{E+}\h{s_2} $ and $\theta\e{E}\h{s_1}$, we would measure the ratio of Einstein radii as 
\begin{align}
    \frac{\theta\e{E+}\h{s_2}}{\theta\e{E}\h{s_1}} &= \frac{(1-\kappa\e{ds_2})(1-\kappa\e{os_1})}{(1-\kappa\e{os_2})(1-\kappa\e{ds_1})}\left(1 + \frac{1-\kappa\e{s_1s_2}}{1-\kappa\e{ds_2}}\frac{\ttheta\e{E,os_1s_2}}{\ttheta\e{E,ods_2}}\right)\eta.
    \label{eq:overestimation}
\end{align}
Thus, unsurprisingly, the mass in s$_1$ can be ignored if and only if 
\begin{equation}
    \ttheta\e{E,os_1s_2} \ll \ttheta\e{E,ods_2}.
    \label{eq:condition}
\end{equation}
For two singular isothermal sphere lenses, this would equate to the assumption that
\begin{equation}
    \frac{D\e{s_1s_2}}{D\e{ds_2}} \ll \frac{\sigma_{v,\mathrm{d}}^2}{\sigma_{v,\mathrm{s}_1}^2},
\end{equation}
which clearly may not always be satisfied.  Of the 4 DSPLs discovered in Euclid Q1 data, one system (The Cosmic Dartboard) has a first source with substantial lensing mass, whereas the other three and the previously known DSPLs can all be treated as a primary lens with a small perturber consistent with \cref{eq:condition}.  

The assumptions of perfect alignment, spherical symmetry and isothermal lenses is of course an oversimplification, and these comments are meant only as a starting point for our intuition. The fundamental point remains, which is that, provided the ratio of angular diameter distances being measured is that in \cref{eq:eta_definition}, the uncertainty will always be a combination of the measurement uncertainty and the line of sight bias in \cref{eq:eta_bias}. If, however, the mass in s$_1$ is not modelled, and only two Einstein radii ($\theta\h{s_2}\e{E+}$ and $\theta\h{s_1}\e{E}$) are observed, then $\eta$ will be systematically overestimated according to \cref{eq:overestimation}. 

\section{Selection effects}
\label{sec:selection_effects}

\begin{figure}
    \centering
    \includegraphics[width=1\textwidth]{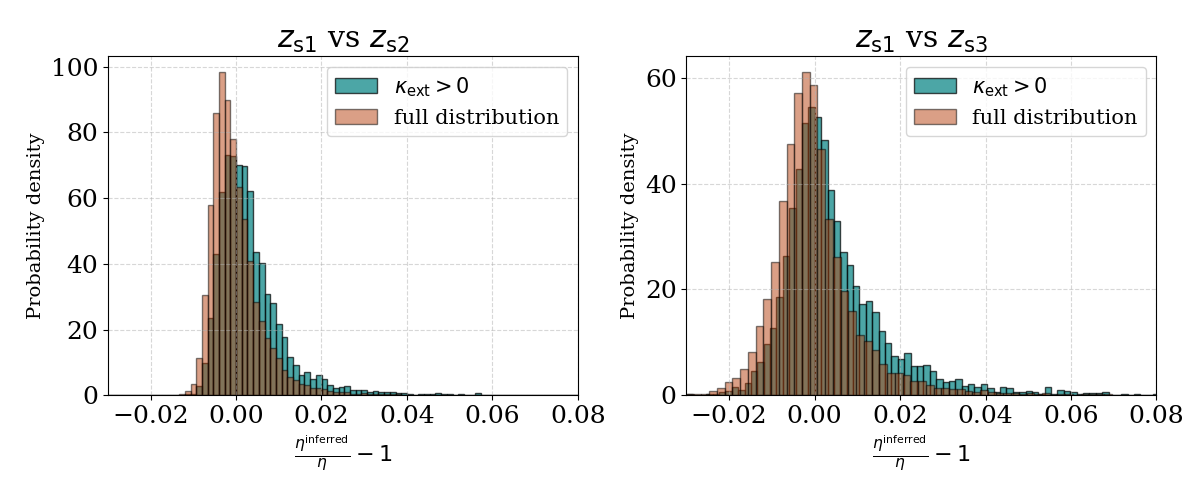}
    \caption{The probability density function for $\eta^\mathrm{inferred}/\eta-1$ for values corresponding to the first and second source planes ($z\e{s_1} = 0.609$, $z\e{s_2} = 2.035$) on the left, and corresponding to the first and third source planes ($z\e{s_1} = 0.609$, $z\e{s_2} = 5.975$) on the right. In orange, we plot the original distributions as in \cref{fig:histograms_jackpot}, and in teal we plot the subset of these distributions when requiring that $\kappa\e{ext}>0$ (as defined relative to the first source). The $\kappa\e{ext}>0$ distribution is shifted towards larger $\eta$ biases by 0.35\% in the $z\e{s1}$ vs $z\e{s2}$ case, and by 1.2\% in the $z\e{s1}$ vs $z\e{s3}$ case.}
    \label{fig:histograms_jackpot_biases}
\end{figure}
In the analysis presented in this paper, we have generally neglected the selection effects coupling specific lines of sight to the presence and detectability of DSPLs. In reality, of course, such couplings exist - lenses are generally found in overdense environments \cite{Keeton_1997}, and with $\kappa\e{ext} \approx \kappa\e{os} + \kappa\e{od} - \kappa\e{ds}$ ($\kappa\e{LOS}$ in the notation of \cite{Fleury_2021a}) larger than 0 \cite{Wells_2024}. Convergence along the line of sight modifies the Einstein radius of a lens, and lenses with lines of sight which result in larger Einstein radii will be both easier to observe and more likely to align with two sources so as to produce a DSPL. 

To gain a sense of the significance of such selection effects, following the conclusions of \cite{Wells_2024}, we can consider the subsample of our simulated lines of sight which correspond to positive $\kappa\e{ext}$ (as defined relative to the first source). \Cref{fig:histograms_jackpot_biases} shows the biases on $\eta$ which we would expect from this subset for the Jackpot lens, compared with the full distributions. The sample corresponding to positive $\kappa\e{ext}$ values, corresponding to around 30\% of all lines of sight, tends to lead to a small bias towards larger $\eta$ values. These lines of sight have median $\kappa\e{ext}$ values of 0.0075 relative to the first source, 0.011 relative to the second, and 0.12 relative to the third, thus broadly consistent with results for the sample of lenses considered in \cite{Wells_2024}. While introducing a small positive bias, this effect is nonetheless subdominant compared to the scatter of the original distributions.

\begin{figure}
    \centering
    \includegraphics[width=1\textwidth]{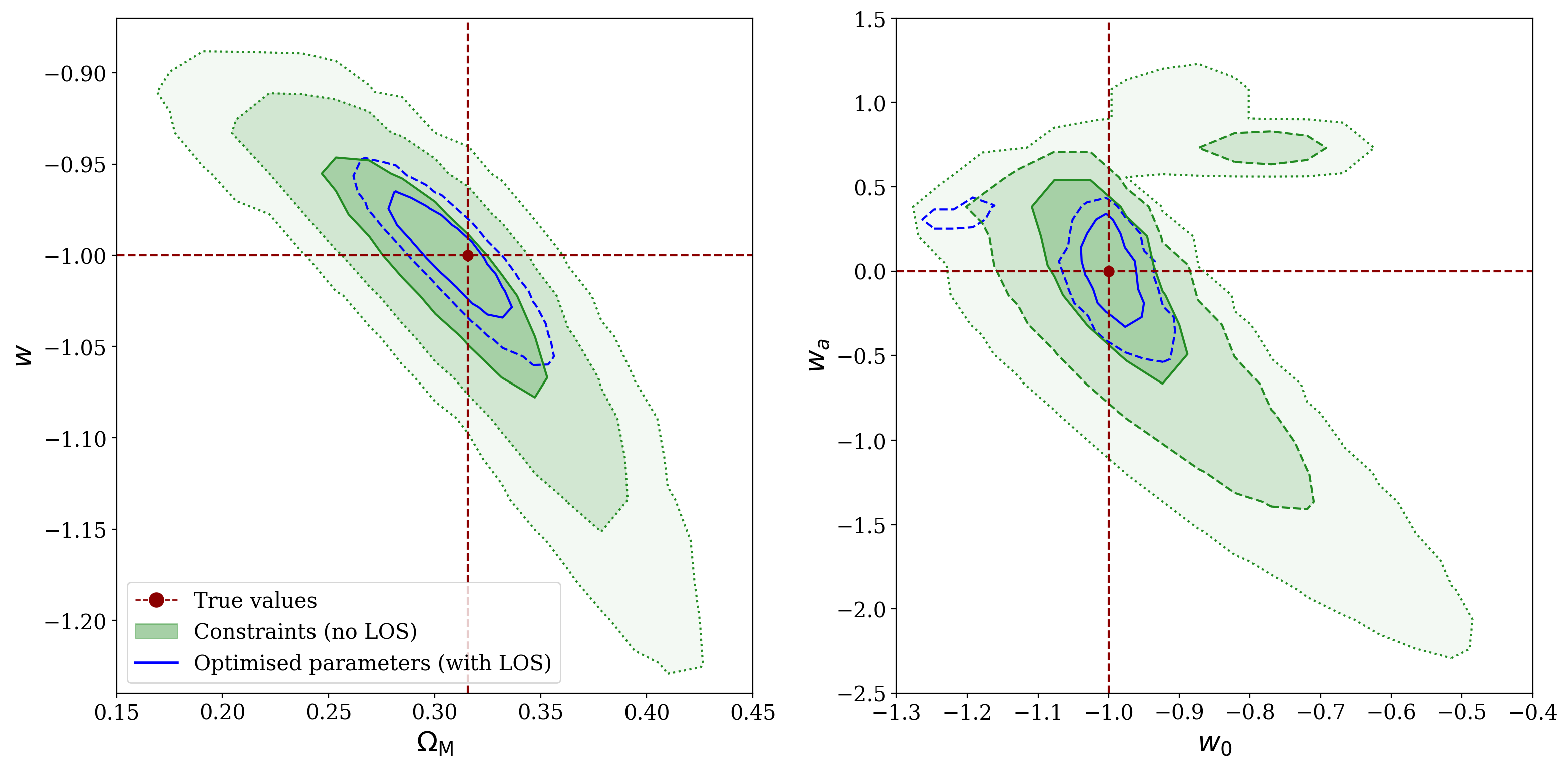} 
    \caption{The scatter introduced by the line of sight compared to the uncertainty in cosmological parameter constraints for a $w$CDM  (left) or $w_0w_a$CDM (right) cosmology, with lines of sight weighted by the factor in \cref{eq:weighting_factor}. In green, we show the 1, 2, and 3-$\sigma$ constraints possible from the forecasted set of 1729 \textit{Euclid} DSPLs and the anticipated measurement uncertainties in $\eta$, with no budget for the line of sight error. In blue, we show the contours for the best guess of these parameters, for 100~000 variations on the lines of sight of each of the DSPLs used in obtaining the constraints. The ``true'' parameter values are plotted in red.}
    \label{fig:weighted_Euclid_cosmology_constraints}
\end{figure}

Key to this paper have been the modifications to Einstein radii by the line of sight. An important selection effect  is the consequence of these changes for the probability of a DSPL occurring. Suppose an observer looks into a solid angle $\Omega$. The probability of two sources being lensed by the same lens is simply the optical depth, i.e. the product of the lensing solid angles as seen by the two sources, and so 
\begin{equation}
    P({\rm s_1, s_2}) = \frac{\Omega\e{DSPL}}{\Omega} = \sum_\ell \frac{\Omega\h{s_1}_\ell}{\Omega}\frac{\Omega\h{s_2}_\ell}{\Omega}.
\end{equation}
If we approximate the lensing cross section as the disk defined by the Einstein radius of the lens with respect to the relevant source, which in turn we take as being the radius within which the average mass density is equal to the critical density, we can write
\begin{equation}
    P({\rm s_1, s_2}) \propto \frac{D\e{ds_1}D\e{ds_2}}{D\e{d}^2D\e{s_1}D\e{s_2}} \, .
    \label{eq:sum_reference}
\end{equation}
Now, nonzero convergence (ie departures from the background density) acts to modify this probability by a factor corresponding to this ratio of angular diameter distances,
\begin{equation}    
    P({\rm s_1, s_2}) \propto \frac{(1-\kappa\e{ds_1})(1-\kappa\e{ds_2})}{(1-\kappa\e{d})^2(1-\kappa\e{s1})(1-\kappa\e{s_2})}.
    \label{eq:weighting_factor}
\end{equation}
As with the effect shown in \cref{fig:histograms_jackpot_biases}, this weighting tends to favour overdense foregrounds and lens planes, and underdense backgrounds. Its effect is best illustrated via its consequences on the cosmological constraints we would obtain from measurements of $\eta$, shown in \cref{fig:weighted_Euclid_cosmology_constraints} for the previously-discussed Euclid forecasts.   

When comparing \cref{fig:weighted_Euclid_cosmology_constraints,fig:Euclid_cosmology_constraints}, we see that, when accounting for the weighting in \cref{eq:weighting_factor}, the centroids of the constraints tend to shift to fractionally higher values of $\Omega\e{m}$, but that no real effect is seen on constraints on the dark energy equation of state. This shift in the $\Omega\e{m}$ centroids is nonetheless significantly smaller than the scatter introduced by the line of sight, and does not represent a systematic overestimation of $\Omega\e{m}$.

A full treatment of the selection effects manifest between lines of sight and DSPLs is beyond the scope of this paper. Nonetheless, from the simple tests run above, it seems that these effects are likely subdominant when compared with the scatter introduced by the line of sight, and the additional uncertainties from this scatter remain a good approximation to the effect of the line of sight.

\bibliographystyle{JHEP.bst}
\bibliography{main.bib}

\end{document}